# `kube-volttron`: Rearchitecting the VOLTTRON Building Energy Management System for Cloud Native Deployment

James Kempf
*UCSC Silicon Valley and Kempf and Associates Consulting*
Santa Clara, CA, USA
kempf42@gmail.com, jkempf@ucsc.edu

*Abstract*—Managing the energy consumption of the built environment is an important source of flexible load and decarbonization, enabling building managers and utilities to schedule consumption to avoid costly demand charges and peak times when carbon emissions from grid generated electricity are highest. A key technology component in building energy management is the building energy management system. Eclipse VOLTTRON is a legacy software platform which enables building energy management. It was developed for the US Department of Energy (DOE) at Pacific Northwest National Labs (PNNL) written in Python and based on a monolithic build-configure-and-run-in-place system architecture that predates cloud native architectural concepts. Yet the software architecture is componentized in a way that anticipates modular containerized applications, with software agents handling functions like data storage, web access, and communication with IoT devices over specific IoT protocols such as BACnet and Modbus. The agents communicate among themselves over a message bus. This paper describes a proof-of-concept prototype to rearchitect VOLTTRON into a collection of microservices suitable for deployment on the Kubernetes cloud native container orchestration platform. The agents are packaged in redistributable containers that perform specific functions and which can be configured when they are deployed. The deployment architecture consists of single Kubernetes cluster containing a central node, nominally in a cloud-based VM, where a microservice containing the database agent (called a "historian") and the web site agent for the service run, and gateway nodes running on sites in buildings where a microservice containing IoT protocol-specific agents handles control and data collection to and from devices, and communication back to the central node.

## I. INTRODUCTION

Buildings constitute over 40% of the primary energy use in developed countries, and over 70% of the electrical load [1]. Yet electrical load is expected to grow by up to 67% over the next 28 years in the US, as the grid quickly decarbonizes and electricity replaces many uses of energy in buildings that today are based on fossil fuels [2]. Clearly, efficiently managing building energy use is necessary to achieve the societal goal of net zero carbon emissions by 2050 while accommodating the increased load. An important technology component required to manage building energy use is a building energy management system. A building energy management system utilizes IoT sensors to detect conditions such as occupancy, temperature, and light levels and conveys that information to a building energy management controller, which then decides when to schedule energy consuming devices to turn on and off. The decisions may be based on a preprogrammed policy or they may derive from a trained machine learning model that incorporates occupant preferences [3]. The building energy management controller can additionally interact with the utility control signals via IEEE 2030.5 [4] or OpenADR [5], sources of weather data, and on-site distributed renewables such as building mounted solar, batteries, and EV chargers to schedule load in a way that reduces costly demand charges and peak loads when carbon emissions from grid electricity are highest. In a 2021 study, the US Department of Energy (DOE) identified such grid-interactive efficient buildings (GEBs) as a potential source of up to 401 terawatt hours of energy savings, 116 gigawatts of peak load savings, for an overall saving of $18 billion and an annual reduction in $CO_2$ emissions on average of 80 million tons [6].

VOLTTRON [7] is an open-source software platform written in Python that has been used to enable a variety of applications, including building energy management. VOLTTRON development started around 2010 at Pacific Northwest National Labs (PNNL) [8] and although PNNL continues to maintain it, in 2016 they engaged with the Eclipse Foundation to bring it under their open-source infrastructure. The VOLTTRON system architecture is based on a build, configure, and run in one place model that is typical of legacy Python applications. This model is quite at variance from today's modular microservices architecture [9], in which applications are developed as a collection of microservices, and distributed as prepackaged containers that are configured when they are deployed onto a container orchestration platform. Yet the agent-based software architecture of VOLTTRON matches the modular container-based deployment model of cloud native applications quite well. The lack of a modular, microservices architecture at the system level may inhibit the ability of organizations to deploy and maintain a building energy management system based on VOLTTRON.

This paper describes `kube-volttron`, a proof-of-concept prototype rearchitecting VOLTTRON from a legacy monolithic application into a microservice based, cloud native application, and implementing the architecture with Docker [10] and Kubernetes [11]. The next section briefly reviews existing work in architecting a building energy management system as a collection of microservices. Section III delves more deeply into the basic VOLTTRON architecture, describing the platform and basic agents that VOLTTRON provides to construct a building energy management service. Section IV describes how the monolithic platform was redesigned to enable configuring VOLTTRON agents as a collection of microservices. In Section V, the Kubernetes deployment is described. Finally, Section VI presents conclusions and suggestions for future work. The contribution of this paper is a demonstration of how to rearchitect a legacy platform for building energy management systems into a collection of modular microservices and deploy the microservices on Kubernetes. The advantages Kubernetes provides to VOLTTRON include easily configurable scalability, load balancing, automatic failover, simpler and more modular upgrade and maintenance, and access to observability tools to enable better service management. The work described in this paper may be useful for system architects and developers who are looking to transition other legacy, monolithic building automation systems into a more loosely coupled collection of microservices deployed onto Kubenetes. The `kube-volttron` code has been published as an open-source repo on Github [12].

## II. REVIEW OF PREVIOUS WORK

Early work on designing a building energy management system as a collection of microservices was primarily conducted in the context of the "smart home". "Smart home" systems can be distinguished from building energy management systems primarily by their additional support for lifestyle-related functions such as, for example, scheduling the lights to be on and music playing when the building occupant returns home from work. Building energy management functions, in contrast, are strictly concerned with managing the building's energy consumption towards the twin goals of reducing expenditure on electricity and reducing carbon emissions.

For example, Bao, et. al. [13] list 10 requirements for a software architecture to support a smart home building operating system. Among the requirements are the need to support the evolution of abstractions, loose coupling between components, and avoiding runtime environment lock-in, all concerns that motivated the work in this paper. However, their primarily focus was on the software architecture and their main concern from a system standpoint is selecting a protocol to interconnect a set of loosely coupled agents. They chose a message bus architecture, a feature that the monolithic VOLTTRON platform already provides. They say nothing about packaging and redistribution nor about deployment and configuration.

More recent work which is similar in nature to `kube-volttron` has focused on building energy management deployed in the "software as a service" (SaaS) paradigm. Haque, et. al. [14] present a hierarchical, three-layer architecture coupled together with APIs based on the HTTP protocol. The layers consist of the Service layer, the API layer, and the Core layer. They contrast a monolithic architecture with a modular architecture, and describe a reimplementation of a modular building energy management system in a microservice-based architecture. According to their classification, VOLTTRON is a modular architecture since it supports modular agents, but they fail to distinguish between the software architecture and system architecture. In addition, the hierarchical nature of their architecture and the use of HTTP APIs instead of a more loosely coupled model can lead to a tighter coupling between microservices. Finally, they don't discuss whether their prototype uses redistributable containers nor whether they deploy onto a cloud native platform.

The work described in [15] is the closest to the `kube-volttron` prototype, in that the authors utilized Kubernetes to deploy containerized application for ingesting, processing, storing, and displaying data from sensor hardware that the authors developed. The data is sent to the Kubernetes framework, called IBFRAME, through an ingress router where it is distributed to multiple services over a pub-sub message bus. The framework supports database services, a web UI, and access to external information such as weather data. All services are packaged as containers and deployed as Kubernetes applications. A major difference between IBFRAME and `kube-volttron` is that IBFRAME only needs to deal with a single type of hardware sensor where the authors controlled the communication protocol. This considerably simplified their data collection architecture.

## III. VOLTTRON ARCHITECTURE

Fig. 1 shows a diagram of the base VOLTTRON deployment architecture. Each node runs a complete set of basic platform agents, depending on which devices and IoT protocols are supported. The central node differs slightly from remote nodes in that it runs an SQL historian agent for storing data in a persistent database, and a VOLTTRON Central agent that supports a dashboard accessed through a build-in web server, in addition to providing routing for local agents and security for remote nodes accessing the platform. A remote node runs the VOLTTRON Central platform agent, which is basically a message bus and RPC router, rather than the VOLTTRON Central agent, and a forwarding historian, which forwards data from the remote node to the central node rather than storing data in a persistent database.

The Message Bus is implemented with either the ZMQ protocol [16] or RabbitMQ [17]. For `kube-volttron`, only ZMQ is supported. VOLTTRON specifies a protocol called the VOLTTRON Interconnect Protocol (VIP) [18] for agents to communicate with each other. VIP consists of two parts: a simple pub-sub format consisting of *topic/subtopic/subtopic* for messages over the message bus and a more complex JSON 2.0

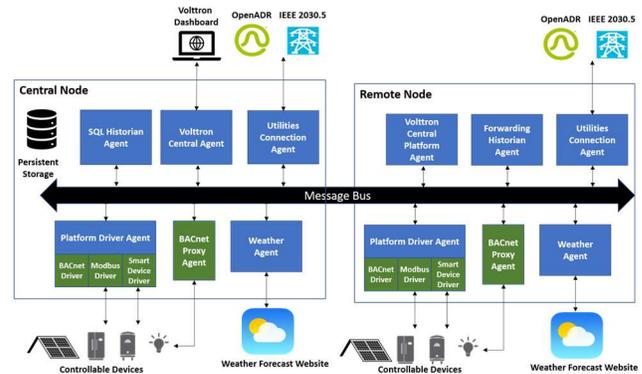

Fig. 1.  VOLTTRON deployment architecture.

HTTP API [19] for point-to-point communication between agents. The JSON 2.0 API defines identity management, authentication, and authorization protocols that agents use to register with the platform and authenticate messages, and is also used by the actuator agent for receiving commands which it conveys to IoT actuator devices.

The platform driver agent is where drivers for various IoT protocols are plugged in. Drivers for protocols used in building energy management systems are available including BACnet [20], Modbus [21], and for smart devices such as the Ecobee wifi API [22]. In most cases, the drivers communicate directly with the devices using the IoT protocol they support or with the manufacturer's web site for smart devices, but for BACnet, a proxy handles communication between the driver and the device. Connections to outside sources of data and control are accomplished through the utility connection agent and the weather agent. The platform supports an SDK for programming new agents in Python. Many agents are available for use cases other than building energy management, for example EV charging or managing utility scale solar farms.

### IV. Converting the VOLTTRON Platform to a Microservice Architecture

In 2021, PNNL released the `volttron-docker` repo on Github. The `volttron-docker` repo consists of the VOLTTRON 8.0 release repackaged with a build and configure Python program, a configuration yaml file, and a Docker build file to build VOLTTRON in a Docker container image. Simple container orchestration is supported through Docker Compose. The containerized VOLTTRON still follows the build-configure-run on the same machine design as the monolithic deployment, the major differences being the yaml file for specifying which agents should be included into the container image and the revamped build process. Container images built with `volttron-docker` are not redistributable outside the node and network in which they have been built, primarily because they use hardcoded IP addresses and port numbers to find other services rather than service discovery.

The first step in converting VOLTTRON to support a microservice architecture was deciding how a service-based deployment would work, ideally with the central node in a public or on-prem cloud and the remote nodes acting as IoT gateways in buildings. This suggested splitting the monolithic deployment into two microservices:

- A microservice containing the SQL historian, persistent database, and VOLTTRON Central agent with web dashboard application and security for deployment in the central node,

- A microservice containing the platform driver agent with drivers for the IoT devices, an actuator agent, a forwarding historian, any agents needed for external access, and the platform agent to connect into the message bus for deployment at remote sites.

The first microservice need not change while the second can be built with customized drivers for the particular IoT protocols, external access agents and the like required at particular sites.

The `volttron-docker` Docker build file was split into two parts:

- A `Dockerfile` that constructs a base image with the operating system (Ubuntu 20.04), python3, the python package manager `pip`, and copies the VOLTTRON code and installs some libraries into the container, then sets up a series of `Dockerfile ONBUILD` commands which are triggered when the microservice is built.

- A `Dockerfile` which specializes the build for a particular microservice type, i.e., for a central node or a remote node. The `ONBUILD` commands trigger when this file is run.

When the microservice is built, the developer must supply four build arguments:

- `VCHOST`: The hostname of the VOLTTRON Central node, defaulting to `vcentral`.

- `VMHOST`: The hostname of the VOLTTRON microservice being built. This also defaults to `vcentral`.

- `VCKEY`: The public key of the VOLTTRON Central node. This is printed out at the end of the build process for the VOLTTRON Central node, if the `VCHOST` and `VMHOST` are the same. The key is required for remote nodes to communicate with the VOLTTRON Central node. This has no default

- `PACKAGES`: A list of additional Python packages that must be installed for the devices. For example, BACnet requires the `bacpypes` package [23]. This also has no default

The lack of service discovery posed the first challenge for converting VOLTTRON to a microservice architecture. DNS hostname service discovery[24], an integral part of the Kubernetes platform, is used by `kube-volttron` to find agents rather than hard coded IP addresses. This allows the microservice container image to be run in different networks and find the services when it boots up. But this required dynamic modification of the VOLTTRON configuration files during the build and preboot phases. During the build process, shell scripts edit the yaml configuration file and configuration files for individual agents to substitute the microservice hostnames for particular patterns corresponding to the service functions. This enables the microservice names to be parameterized according to, for example, the location of the remote site. When the URL for the service is accessed at run time, the hostname is resolved to its address in the DNS A record.

While the ZMQ API supports DNS resolution for the peer address, DNS resolution does not work for the VIP bus address. Since the VOLTTRON platform code does not have a solid abstraction layer for the message bus (for example, clients need to be aware of whether RabbitMQ or ZMQ is used), a preboot Python program resolves the VIP bus URL host names and edits them into the configuration files prior to starting the VOLTTRON microservice.

The result of these changes is to split the build-configure-run script into a collection of two shell scripts and a Python program that run when microservice image is built and a shell script and Python program that run when the microservice boots. OpenSSL

```yaml
apiVersion: "k8s.cni.cncf.io/v1"
kind: NetworkAttachmentDefinition
metadata:
  name: bacnet
spec:
  config: '{
      "cniVersion": "0.4.0",
      "name": "bacnet",
      "type": "host-device",
      "device": "enp0s8",
      "ipam": {
        "type": "dhcp"
      }
    }'
```

Fig. 2  NetworkAttachmentDefinition yaml manifest

[25] was substituted for certificate configuration rather than the built-in Python-based configuration provided with `volttron-docker`, to allow provisioning of certificates and installation in the container by the developer rather than creating them on the fly. These changes were developed in the `microservice-volttron` Github repo.

## V. Deploying the Microservices to Kubernetes

For secure communication between the VOLTTRON Central node and the gateway nodes, a Wireguard [26] point to point VPN is used. Wireguard is a modern, kernel-based VPN that is easy to configure and lightweight on the wire. The `vcentral` microservice is deployed on the central node and one of two IoT protocol services (in the prototype) are deployed on the gateway node: `vremote`, a microservice configured with simulated IoT device that comes as part of the VOLTTRON distribution, and `vbac`, a microservice, with the BACnet driver and proxy agent configured. The former is just for testing, the latter can be used with a simulated BACnet device that runs as a Python program to demonstrate BACnet connectivity into the site network.

For BACnet connectivity, the primary technical challenge was how to open a second interface between the BACnet proxy running in the `vbac` pod and the local site network running on the host. The second interface is required because BACnet/IP uses broadcast for device discovery and the VOLTTRON BACnet proxy needs to be able to directly connect with devices using their IP addresses. The Flannel [27] container network interface (CNI) plug-in is used by `kube-volttron` as the intra-cluster, pod-to-pod network provider, but Flannel prohibits direct access to the host's site network for security. It constructs an IP overlay network using VXLAN at a fixed CIDR (10.244.0.0/24) for intra-cluster networking. To enable a second interface, the Multus [28] multi-interface driver was used. Multus supports one interface on the Flannel intra-cluster network and allows others to be configured.

The Kubernetes CNI [29] provides a custom resource definition (CRD) called a `NetworkAttachmentDefinition` for declaring the type of the second interface. The yaml manifest for the `vbac` microservice second interface, named `bacnet`, is shown above in Fig. 2. The manifest includes an embedded JSON object with the interface configuration. The `bacnet` interface is of `host-local`, type which means that it absorbs a second interface from the host into the pod's network namespace when the pod is created. IP address management (ipam) is provided by a DHCP relay running on the host which relays IP addresses from the local subnet DHCP server. Fig. 3 below illustrates the network architecture of a gateway node running the `vbac` BACnet microservice.

For each VOLTTRON microservice, a Kubernetes a `Deployment` type with one replica to deploy the pod is defined. As part of the `vbac` Deployment manifest, the pod spec includes a property declaring that the `bacnet` interface definition should be used to construct the second interface. The `Service` type defines two ports for the VIP protocol, 8443 for the HTTP protocol and 22916 for the ZMQ message bus protocol. A `Service` type is also defined for each microservice to decouple the actual pod IP address from the IP address at which it is accessed, so if the pod is upgraded or crashes, clients see minimal disruption in service. In addition, the `vcentral` microservice declares an external IP address - the address of an external interface on the VM or host where it is running – to access the web dashboard app. While this method of exposing a web service in Kubernetes is not recommended, the other methods all restrict the ports that can be exposed or require the deployment of additional services like a load balancer.

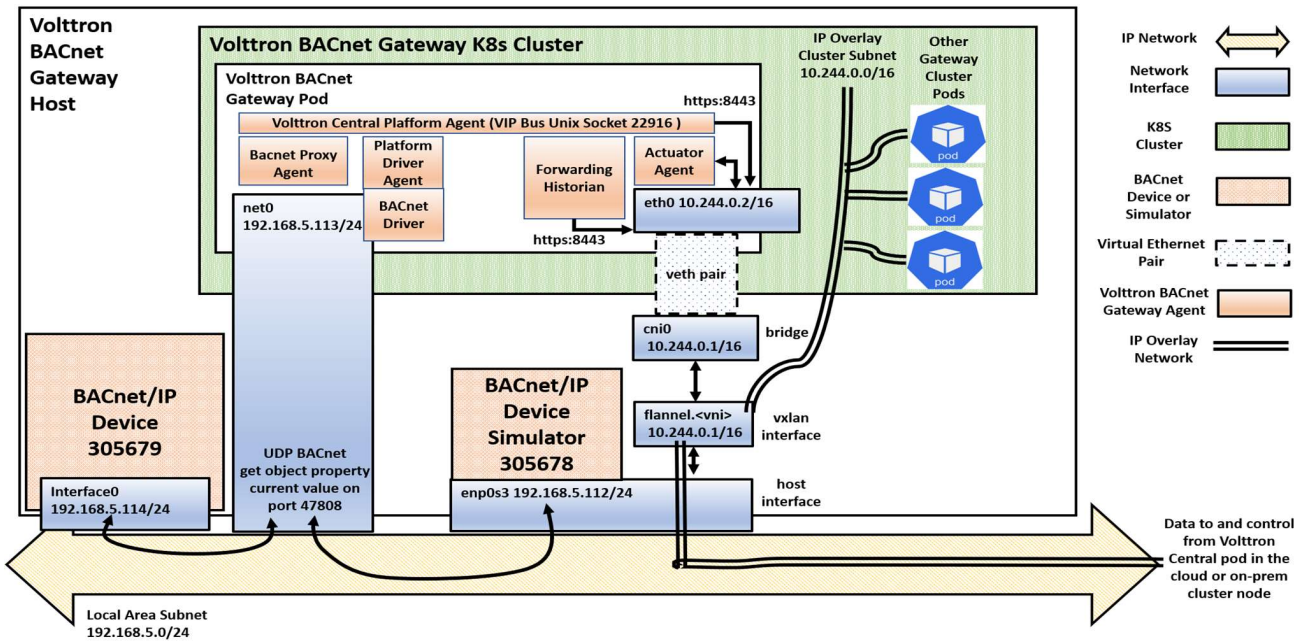

Fig. 3  Network architecture of the gateway node pod with BACnet microservice `vbac`

The `kube-volttron` implementation uses two `Configmap` type objects on the gateway node to configure the `vbac` microservice with the IP address of the BACnet device when the microservice is deployed, since VOLTTRON does not yet use broadcast for on-the-fly device discovery. The devops engineer simply edits the `Configmaps` and inserts the IP address of the BACnet device on the local subnet before deploying the microservice. The `kube-volttron` Github repo includes a small Python program that simulates an air handling unit for demoing `vbac`.

On the central node, the Historian database file is mounted from the local node file system using a `PersistentVolume` of type `local`. This allows the database to survive if the microservice pod crashes or is upgraded. The `vcentral` microservice uses the SQLite database which is the VOLTTRON default. Some experiments were done using a Kubernetes installable version of Postgres, but since this also mounts storage from the local file system, the resilience and reliability properties were similar to SQLite, so it was abandoned.

## VI. CONCLUSIONS AND FUTURE WORK

During the development of `microservice-volttron` and `kube-volttron`, an attempt was made to minimize any changes to the VOLTTRON code itself and just work with the deployment code in `volttron-docker`, primarily because PNNL is in the process of rearchitecting the base platform to be more modular for the Version 10.0 release. For this release, the code is being broken up into functional components to enable more focused deployments while keeping scalability and usability in mind.

There are some components of VOLTTRON which could be replaced by open-source components without substantially modifying the basic VOLTTRON design. Many of these changes would benefit deployment on platforms other than Kubernetes as well.

As noted above, VOLTTRON currently uses hard coded IP addresses and port numbers to identify services. VOLTTRON services, including the VIP message bus service, could use DNS service discovery, with the service name being either the hostname of the service as `kube-volttron` does or by recording the service name, IP address, and port number in a DNS SRV record [35]. Using the DNS SRV record would have the added advantage of removing the need to specify the ports in the VOLTTRON service configuration URLs. The configuration URLs would only need to specify the protocol and the service name. For `kube-volttron`, using DNS service discovery would remove a substantial chunk of build time and preboot time code.

Another area where VOLTTRON could benefit from incorporating more recent open-source components is identity and access management (IAM), both for workloads and for users. The bespoke JSON API for workload identity management, authentication, and authorization could be replaced with. SPIFFE [31]. SPIFFE uses X.509 certificates for identifying workloads and is a general workload IAM mechanism not tied exclusively to Kubernetes. Kubernetes itself has well designed and tested mechanisms for handling cryptographic material and supports a `Secrets` object type [32] for storing encrypted passwords and private keys. `Secrets` objects could be used to enable service accounts in VOLTTRON, which grant fine-grained role-based access control to particular services. User IAM through the dashboard could be modified to use the OAuth [33] and OIDC [34]. These are also general protocols not tied exclusively to Kubernetes, and would remove the continued need to maintain the user IAM code in the VOLTTRON platform itself.

Further enhancements may require deeper changes in the VOLTTRON platform design. For example, the combined HTTP JSON API and message bus architecture could be replaced with a service mesh [36]. A service mesh replaces the communication support built into the application or provided by a library component with a side car container, that handles communication between applications, including security such as authentication and encryption with mutual TLS. A service mesh splits communication into a control plane for monitoring and configuring communications, and a data plane over which the applications exchange data, and would open VOLTTRON to cloud-native observability solutions for service performance monitoring.

An area of exploration for `kube-volttron` is using a flat network CNI plug-in rather than Flannel. Flat network plug-ins such as Calico [37] and Cilium [37] deploy the intra-cluster network in the same CIDR range as the host. This would remove the need for Multus and the second interface on the `vbac` microservice pod. The disadvantage is that management of a flat intra-cluster network requires more sophisticated policy mechanisms for security. Flannel depends on network isolation provided by the overlay for security. Any traffic admitted into the Flannel overlay comes through a Kubernetes proxy. While the second interface increases also the cluster attack surface, the increase is bounded and if BACnet is not used in a particular deployment, the second interface need not be deployed.

Finally, `microservice-volttron` and `kube-volttron` have not fully explored the potential of a microservice architecture for VOLTTRON. The `vcentral`, `vbac`, and `vremote` microservices bundle several agents into one container. Reducing the granularity down to a single agent would enable easier, more modular upgrade, and maintenance. In addition, services that take advantage of open-source libraries written in other languages or are implemented in other languages than Python could be incorporated into a VOLTTRON deployment.

In conclusion, although VOLTTRON has a long history and, as with many legacy software platforms, has accumulated some technical debt, it nevertheless has a software architecture that, with a modest amount of additional work on its system architecture, could extend it to a cloud native deployment. Such work could substantially enhance VOLTTRON's ability to contribute to decarbonizing societal energy systems by providing a solid open-source platform on which building energy management service providers could build their services.

ACKNOWLEDGMENT

The author would like to thank Ben Barting for providing his air handling unit simulator in a public Github repo, Tanya Barham of Community Energy Labs, and John Powers of ExtensibleEnergy for introducing him to the importance of a building energy management systems for building decarbonization, and the folks at PNNL for their continued work on VOLTTRON over the years. Special thanks are also due to Jereme Haack and Shwetha Niddodi of PNNL for providing extensive feedback on the paper.